\documentclass[12pt,twoside]{article}

\usepackage{amsmath}
\usepackage{amssymb}
\usepackage{cite}
\usepackage{ccaption}
\usepackage{hyperref}
\usepackage{graphicx}
\captiondelim{.}

\begin{document}
\newcommand{\pst}{\hspace*{1.5em}}

\newcommand{\rigmark}{\em Journal of Russian Laser Research}
\newcommand{\lemark}{\em Volume 30, Number 5, 2009}

\newcommand{\be}{\begin{equation}}
\newcommand{\ee}{\end{equation}}
\newcommand{\bm}{\boldmath}
\newcommand{\ds}{\displaystyle}
\newcommand{\bea}{\begin{eqnarray}}
\newcommand{\eea}{\end{eqnarray}}
\newcommand{\ba}{\begin{array}}
\newcommand{\ea}{\end{array}}
\newcommand{\arcsinh}{\mathop{\rm arcsinh}\nolimits}
\newcommand{\arctanh}{\mathop{\rm arctanh}\nolimits}
\newcommand{\bc}{\begin{center}}
\newcommand{\ec}{\end{center}}

\thispagestyle{plain}

\label{sh}


\begin{center} {\Large \bf
\begin{tabular}{c}
Virtual correlations in single qutrit
\end{tabular}
} \end{center}

\bigskip

\bigskip

\begin{center} {\bf
A. A. Strakhov$^1$ and V. I. Man'ko$^{1,2}$
}\end{center}

\medskip

\begin{center}
{\it
$^1$Moscow Institute of Physics and Technolodgy\\
Institutskii per. 9, Dolgoprudnyi, Moscow Region 141700, Russia

\smallskip

$^2$P.N. Lebedev Physical Institute, Russian Academy of Sciences\\
Leninskii Prospect 53, Moscow 119991, Russia
}
\smallskip

\end{center}

\begin{abstract}\noindent
We construct the positive invertible map of the mixed states of a single qutrit onto the states of two identical fermions. It is shown that using this one-to-one correspondence between qutrit states and states of two identical three-dimensional fermions (the unphysical odd dimension is picked for simplicity) one may attribute hidden entanglement to a single mixed state of qutrit.
\end{abstract}

\medskip


\section{Introduction}
\pst
Quantum entanglement~\cite{1} is one of the most valueable nonclassical resources in quantum information science. By virtue of entanglement the protocols of superdense coding~\cite{2} and quantum teleportation~\cite{3} were discovered. Entangled states were proved to be necessary in Shor's and Grover's algorithms~\cite{4,5} and are used in various cryptography protocols~\cite{6,7}. Entangled states are also used in more applied areas of physics such as quantum metrology~\cite{8,9}. These are few of the whole list of entanglement applications.

Naturally it is hard to prepare and maintain maximally entangled states. In order to control the amount of entanglement in quantum systems several measures were proposed. In case of bipartite pure states a well-defined measure was proposed in~\cite{10}. The case of multiparticle pure states~\cite{11} and mixed states~\cite{12} is much more complicated. In particular for mixed states there were introduced the following measures: the entanglement of formation~\cite{12}, the distillable entanglement~\cite{13}, the relative entropy of entanglement~\cite{14}, the Negativity and logarithmic Negativity~\cite{15,16}. The last of them, i. e. Negativity, has an advantage that it can be simply calculated. We will use it further in the article.

Recently there were conducted several attempts to study quantum correlations of noncomposite qudit systems~\cite{17,18,19,20,21}. The physical realization of such systems is convenient with nonlinear quantum circuits on Josephson junctions~\cite{22,23}. This paper is an attempt to study correlations between different degrees of freedom in qutrit via the corresponding bipartite system of indistinguishable fermions. The source of the correlations is an entanglement between these fermions provided by antisymmetry. It is worth noting that entanglement in systems of indistinguishable particles was studied as in general case~\cite{24}, as in case of not relying on (anti)symmetry of their states~\cite{25}. Another interesting insight is connecting the degree of entanglement in composite bosons and fermions with deformation parameters of the corresponding deformed oscillator algebras~\cite{26,27}.

\section{Partial trace in matrix representation}
\pst
The state of a subsystem of a quantum system is defined by an operation called partial trace - taking mean value over states of another subsystem:

\be
\label{partialtrace}
\rho_\mathcal{I} = \text{Tr}_{\mathcal{J} \setminus \mathcal{I}}[\rho_{\mathcal{J}}]\Leftrightarrow \langle i_1|\rho_\mathcal{I}|i_2\rangle=\sum\limits_{|j\rangle\in\mathcal{J} \setminus \mathcal{I}}\langle i_1|\langle j|\rho_{\mathcal{J}}|j\rangle|i_2\rangle,
\ee
where $|i_1\rangle$ and $|i_2\rangle$ are basis vectors of the subsystem $\mathcal{I}$ and $|j\rangle$ - basis vectors of the subsystem $\mathcal{J} \setminus \mathcal{I}$. Despite the explicit choice of basis vectors of space $\mathcal{J}$, partial trace is invariant with respect to unitary transformations of $\mathcal{I}$.

Let us consider a space $\mathcal{J}=H_3 \otimes H_2$ and choose a lexicographically ordered basis $|e_1\rangle=|a_1\rangle\otimes|b_1\rangle,\;|e_2\rangle=|a_1\rangle\otimes|b_2\rangle,\ldots,|e_6\rangle=|a_3\rangle\otimes|b_2\rangle$. Then an arbitrary state $\rho$ will be presented as matrix:

\be
\rho=
\begin{pmatrix}
\rho_{11} & \rho_{12} & \rho_{13} & \rho_{14} & \rho_{15} & \rho_{16} \\
\rho_{21} & \rho_{22} & \rho_{23} & \rho_{24} & \rho_{25} & \rho_{26} \\
\rho_{31} & \rho_{32} & \rho_{33} & \rho_{34} & \rho_{35} & \rho_{36} \\
\rho_{41} & \rho_{42} & \rho_{43} & \rho_{44} & \rho_{45} & \rho_{46} \\
\rho_{51} & \rho_{52} & \rho_{53} & \rho_{54} & \rho_{55} & \rho_{56} \\
\rho_{61} & \rho_{62} & \rho_{63} & \rho_{64} & \rho_{65} & \rho_{66}
\end{pmatrix}
\ee
Formula (\ref{partialtrace}) then might be represented by square block matrices which sizes are determined by the numbers of dimensions of spaces $H_3$ and $H_2$:

\be
\rho=
\left(
\begin{array}{cc|cc|cc}
\rho_{11} & \rho_{12} & \rho_{13} & \rho_{14} & \rho_{15} & \rho_{16} \\
\rho_{21} & \rho_{22} & \rho_{23} & \rho_{24} & \rho_{25} & \rho_{26} \\
\hline
\rho_{31} & \rho_{32} & \rho_{33} & \rho_{34} & \rho_{35} & \rho_{36} \\
\rho_{41} & \rho_{42} & \rho_{43} & \rho_{44} & \rho_{45} & \rho_{46} \\
\hline
\rho_{51} & \rho_{52} & \rho_{53} & \rho_{54} & \rho_{55} & \rho_{56} \\
\rho_{61} & \rho_{62} & \rho_{63} & \rho_{64} & \rho_{65} & \rho_{66}
\end{array}
\right)
=
\begin{pmatrix}
R_{11} & R_{12} & R_{13} \\
R_{21} & R_{22} & R_{23} \\
R_{31} & R_{32} & R_{33}
\end{pmatrix};
\ee
\be
\rho_1=
\begin{pmatrix}
\text{Tr}R_{11} & \text{Tr}R_{12} & \text{Tr}R_{13} \\
\text{Tr}R_{21} & \text{Tr}R_{22} & \text{Tr}R_{23} \\
\text{Tr}R_{31} & \text{Tr}R_{32} & \text{Tr}R_{33}
\end{pmatrix},
\quad
\rho_2=
R_{11}+R_{22}+R_{33},
\ee
where $\rho_1$ is a state of the first subsystem $H_3$ and $\rho_2$ - a state of the second subsystem $H_2$. Partial traces in matrix form will be constantly used further in this paper.

\section{Qutrit as a $\mathcal{H}=H_3\wedge H_3$ fermionic system.}
\pst
Let us pick a basis $\{|e_1\rangle$, $|e_2\rangle$, $|e_3\rangle\}$ in an arbitrary three dimensional state space $\mathcal{Q}$. The general mixed state of qutrit is described by a density operator $\rho_I\in S(\mathcal{Q})$: $\rho_I=\sum\limits_{i,j=1}^3(\rho_I)_{ij}|e_i\rangle\langle e_j|,$ in matrix form:
\be
\rho_I=
\begin{pmatrix}
\rho_{11} & \rho_{12} & \rho_{13} \\
\rho_{21} & \rho_{22} & \rho_{23} \\
\rho_{31} & \rho_{32} & \rho_{33} \\
\end{pmatrix}
\ee

Now we will consider a system of two identical three-dimensional fermions. Despite fermion state dimension is odd which is unphysical it is picked so for simplicity. Mathematically it is an exterior product of two three-dimensional spaces which means that it is antisymmetric with respect to permutations: $\mathcal{H}=H_3\wedge H_3$. Let us denote basis vectors in $H_3$ as $|1\rangle,|2\rangle,|3\rangle$. Then in the basis of exterior product an arbitrary pure state $|\Psi\rangle\in\mathcal{H}$ looks as follows:

\be
|\Psi\rangle=a_1 |g_1\rangle+a_2 |g_2\rangle+a_3 |g_3\rangle,
\label{Psi1}
\ee
$$
|g_1\rangle=\frac {1}{\sqrt{2}}|1\rangle\wedge|2\rangle=\frac {1}{\sqrt{2}}(|1\rangle\otimes|2\rangle-|2\rangle\otimes|1\rangle),
$$
\be
|g_2\rangle=\frac {1}{\sqrt{2}}|1\rangle\wedge|3\rangle=\frac {1}{\sqrt{2}}(|1\rangle\otimes|3\rangle-|3\rangle\otimes|1\rangle),
\ee
$$
|g_3\rangle=\frac {1}{\sqrt{2}}|2\rangle\wedge|3\rangle=\frac {1}{\sqrt{2}}(|2\rangle\otimes|3\rangle-|3\rangle\otimes|2\rangle),
$$

We will further omit the $\otimes$ symbol of tensor product. Thus we see that the space $\mathcal{H}$ is isomorphic to qutrit space $\mathcal{Q}$. The general form of density operator $\rho_O\in S(\mathcal{H})$ on space $\mathcal{H}$ is: $\rho_O=\sum\limits_{i,j=1}^3(\rho_O)_{ij}|g_i\rangle\langle g_j|$, so we can introduce an isomorphism between sets of mixed states $S(\mathcal{Q})$ and $S(\mathcal{H})$ by equalizing matrix elements $(\rho_I)_{ij}=(\rho_O)_{ij}=\rho_{ij}$ for all $i,j=1,2,3$.

As we already mentioned, $\rho_O$ is a state in antisymmetrized subspace of the tensor product space $H_3\otimes H_3$. Let us observe this state in context of the whole space $H_3\otimes H_3$. It is easy to verify that its matrix representation is as follows:
\be
\rho'_O=\frac {1}{2}\begin{pmatrix}
0 & 0 & 0 & 0 & 0 & 0 & 0 & 0 & 0 \\
0 & \rho_{11} & \rho_{12} & -\rho_{11} & 0 & \rho_{13} & -\rho_{12} & -\rho_{13} & 0 \\
0 & \rho_{21} & \rho_{22} & -\rho_{21} & 0 & \rho_{23} & -\rho_{22} & -\rho_{23} & 0 \\
0 & -\rho_{11} & -\rho_{12} & \rho_{11} & 0 & -\rho_{13} & \rho_{12} & \rho_{13} & 0 \\
0 & 0 & 0 & 0 & 0 & 0 & 0 & 0 & 0 \\
0 & \rho_{31} & \rho_{32} & -\rho_{31} & 0 & \rho_{33} & -\rho_{32} & -\rho_{33} & 0 \\
0 & -\rho_{21} & -\rho_{22} & \rho_{21} & 0 & -\rho_{23} & \rho_{22} & \rho_{23} & 0 \\
0 & -\rho_{31} & -\rho_{32} & \rho_{31} & 0 & -\rho_{33} & \rho_{32} & \rho_{33} & 0 \\
0 & 0 & 0 & 0 & 0 & 0 & 0 & 0 & 0 \\
\end{pmatrix}.
\ee
Now using the partial trace operation, introduced in the previous section, we are able to obtain single three-dimensional fermionic states:
\be
\rho_1=\rho_2=\frac {1}{2}\begin{pmatrix}
\rho_{11}+\rho_{22} & \rho_{23} & -\rho_{13} \\
\rho_{32} & \rho_{11}+\rho_{33} & \rho_{12} \\
-\rho_{31} & \rho_{21} & \rho_{22}+\rho_{33} \\
\end{pmatrix}.
\ee
The map $\mathcal{M}:\rho_I\mapsto \rho_{1,2}$ is a completely positive trace-preserving (CPTP) map. It is obvious because the map $\rho_I\mapsto \rho_O$ is an isomorphism and expanding $\rho_O$ to $\rho'_O$ is simply adding extra dimensions to Hilbert space which do not participate in constructing of the state $\rho'_O$, i.e. the spectrum of $\rho'_O$ is equal to the spectrum of $\rho_O$ and hence of $\rho_I$. The partial trace is obviously a CPTP map, so the whole map $\mathcal{M}$ is CPTP. Despite of such trivial construction of reduced states $\rho_{1,2}$ in general their spectrum is different from $\rho_I$ because their construction was out of fermionic antisymmetric subspace $\mathcal{H}\subset H_3\otimes H_3$.

\section{Correlations between different degrees of freedom of qutrit}
\pst
The well-known Peres-Horodecki criterion, or PPT criterion, is a necessary condition for a mixed state to be separable. It states that if the density operator $\rho$ of a bipartite system is separable then partial transpose of one of the subsystems is positive. The partial transpose of the second subsystem of $\rho'_O$ in matrix form is the following:
\be
\rho_O^{'PT}=\frac {1}{2}\begin{pmatrix}
0 & 0 & 0 & 0 & -\rho_{11} & -\rho_{21} & 0 & -\rho_{12} & -\rho_{22} \\
0 & \rho_{11} & \rho_{21} & 0 & 0 & 0 & 0 & -\rho_{13} & -\rho_{23} \\
0 & \rho_{12} & \rho_{22} & 0 & \rho_{13} & \rho_{23} & 0 & 0 & 0 \\
0 & 0 & 0 & \rho_{11} & 0 & -\rho_{31} & \rho_{12} & 0 & -\rho_{32} \\
-\rho_{11} & 0 & \rho_{31} & 0 & 0 & 0 & \rho_{13} & 0 & -\rho_{33} \\
-\rho_{12} & 0 & \rho_{32} & -\rho_{13} & 0 & \rho_{33} & 0 & 0 & 0 \\
0 & 0 & 0 & \rho_{21} & \rho_{31} & 0 & \rho_{22} & \rho_{32} & 0 \\
-\rho_{21} & -\rho_{31} & 0 & 0 & 0 & 0 & \rho_{23} & \rho_{33} & 0 \\
-\rho_{22} & -\rho_{32} & 0 & -\rho_{23} & -\rho_{33} & 0 & 0 & 0 & 0 \\
\end{pmatrix}.
\ee
It can be easily verified that $\rho_O^{'PT}$ has negative values for various $\rho_I$ states, i.e. it's an entangled state. Actually, with the aid of entanglement monotone called Negativity it's possible to show that this state is always entangled. This has a rather clear physical meaning - combining two qutrits and restricting them to be identical fermions we freeze their degrees of freedom which correspond to the maximal total spin of the combined system. Thus such states are necessarily entangled.

The entanglement monotone $E$ is nonnegative linear functional over state $\rho$ with two main properties:

1) The monotone is a convex functional: $E(\sum\limits_i p_i\rho_i)\leq\sum\limits_i p_i E(\rho_i)$.

2) The functional is an entanglement monotone, i.e. it doesn't increase on average under local quantum operations and classical communication (LOCC).

Most of the entanglement monotones, like distillability monotones and monotones of formation, are immensely difficult to compute. The more easily countable measure of entanglement is called Negativity monotone. It is defined as follows:
\be
E(\rho)=\frac {\|\rho^{PT}\|_1-1}{2}=\sum\limits_i \frac {|\lambda_i|-\lambda_i}{2},
\ee
where $\|X\|_1=\text{Tr}(\sqrt{X^{\dag}X})$ is a trace norm. In case of Hermitian operators it's just a sum of modules of negative eigenvalues.

Now it is possible to show that in case of initial pure states $\rho_I=|\psi\rangle\langle\psi|$ their Negativity is $E=\frac {1}{2}$ and in case of maximally chaotic state $\rho_I=\begin{pmatrix}\frac {1}{3} & 0 & 0 \\ 0 & \frac {1}{3} & 0 \\ 0 & 0 & \frac {1}{3} \end{pmatrix}$ $E=\frac {1}{3}$. More formally, one can use another more convenient entanglement monotone for pure states $|\psi\rangle\in H_A\otimes H_B$: the entropy of the subsystem $S=S_A=S_B=-\text{Tr}\rho_A\ln\rho_A.$ In our case if the initial state $\rho_{I}=|\psi\rangle\langle\psi|$ and $|\psi\rangle=a_1|e_1\rangle+a_2|e_2\rangle+a_3|e_3\rangle$ then the density matrices of fermion subspaces:
\be
\rho_1=\rho_2=\frac {1}{2}\begin{pmatrix}
|a_1|^2+|a_2|^2 & a_2a^*_3 & -a_1a^*_3 \\
a^*_2a_3 & |a_1|^2+|a_3|^2 & a_1a^*_2 \\
-a^*_1a_3 & a^*_1a_2 & |a_2|^2+|a_3|^2
\end{pmatrix}.
\ee
A simple technical calculation shows that nonzero eigenvalues of this matrix are $\lambda_{1,2}=\frac {1}{2}.$ This means that the entanglement measure of $\rho'_{O}$ is $S=\ln 2$ for all initial pure states $\rho_{I}$. Thus the Negativities $E$ of all pure states $\rho'_{O}$ are also equal to each other. En elementary calculation for an arbitrary pure state shows that $E=\frac {1}{2}$.

The previous considerations about pure states Negativities prove that for two different mixed states with equal coefficients $\rho_{I1}=p_1|\psi_1\rangle\langle\psi_1|+p_2|\psi_2\rangle\langle\psi_2|+p_3|\psi_3\rangle\langle\psi_3|$ and $\rho_{I2}=p_1|\phi_1\rangle\langle\phi_1|+p_2|\phi_2\rangle\langle\phi_2|+p_3|\phi_3\rangle\langle\phi_3|$ Negativities are equal to each other. Thus to analyze all mixed states it is enough to consider a general density matrix:
\be
\rho_{I}=\begin{pmatrix}
p_1 & 0 & 0 \\
0 & p_2 & 0 \\
0 & 0 & p_3
\end{pmatrix}.
\ee
The partial transpose of $\rho'_{O}$ then will be the following:
\be
\rho_O^{'PT}=\frac {1}{2}\begin{pmatrix}
0 & 0 & 0 & 0 & -p_1 & 0 & 0 & 0 & -p_2 \\
0 & p_1 & 0 & 0 & 0 & 0 & 0 & 0 & 0 \\
0 & 0 & p_2 & 0 & 0 & 0 & 0 & 0 & 0 \\
0 & 0 & 0 & p_1 & 0 & 0 & 0 & 0 & 0 \\
-p_1 & 0 & 0 & 0 & 0 & 0 & 0 & 0 & -p_3 \\
0 & 0 & 0 & 0 & 0 & p_3 & 0 & 0 & 0 \\
0 & 0 & 0 & 0 & 0 & 0 & p_2 & 0 & 0 \\
0 & 0 & 0 & 0 & 0 & 0 & 0 & p_3 & 0 \\
-p_2 & 0 & 0 & 0 & -p_3 & 0 & 0 & 0 & 0 \\
\end{pmatrix}.
\ee
Its negative eigenvalues are the negative solutions of the third order equation
\be
\lambda^3-(p_1^2+p_2^2+p_3^2)\lambda+2p_1p_2p_3=0.
\ee
Careful analysis shows that for every distribution $p_i$ there is only one negative solution. The infinum of Negativity is situated at maximally chaotic state $p_1=p_2=p_3=\frac {1}{3}$ and is equal to $E=\frac {1}{3}$.

\section{Conclusions}
\pst
In this paper the simplest nontrivial case of quantum correlations in qutrit was analyzed. It was demonstrated that for arbitrary mixed qutrit state there exist nonzero Negativity for corresponding virtual bipartite system of identical three-dimensional fermions. Thus the correlations between the degrees of freedom of qutrit emerge. These considerations could be generalized for qudits with more degrees of freedom in a straightforward manner.

\end{document}